\documentclass[aps,prl,twocolumn,showpacs,preprintnumbers,amsmath,amssymb,superscriptaddress,floatfix]{revtex4} 
\usepackage{graphicx}
\usepackage{amssymb}
\usepackage{epstopdf}
\usepackage{bm}

\newcommand*{\m}[0]{\ensuremath{\mathbf{m}}}
\newcommand*{\p}[0]{\ensuremath{\mathbf{p}}}
\newcommand*{\vp}[0]{\ensuremath{\varphi}}
\newcommand*{\dd}[0]{\ensuremath{\text{d}}}

\newcommand*{\deriv}[2]{\ensuremath{\frac{\text{d} #1}{\text{d} #2}}}

\begin{document}
\title{Time Domain Mapping of Spin Torque Oscillator Effective Energy}
\author{Graham E. Rowlands}
\affiliation{Department of Physics and Astronomy, University of California, Irvine, California 92697, USA}
\author{Jordan A. Katine}
\affiliation{Hitachi Global Storage Technologies, 3403 Yerba Buena Road, San Jose, CA 95135}
\author{Juergen Langer}
\affiliation{Singulus Technologies, 63796 Kahl am Main, Germany}
\author{Jian Zhu}
\affiliation{Department of Physics and Astronomy, University of California, Irvine, California 92697, USA}
\author{Ilya N. Krivorotov}
\affiliation{Department of Physics and Astronomy, University of California, Irvine, California 92697, USA}
\date{\today}
\begin{abstract}
Stochastic dynamics of spin torque oscillators (STOs) can be described in terms of magnetization drift and diffusion over a current-dependent effective energy surface given by the Fokker-Planck equation. Here we present a method that directly probes this effective energy surface via time-resolved measurements of the microwave voltage generated by a STO. We show that the effective energy approach provides a simple recipe for predicting spectral line widths and line shapes near the generation threshold. Our time domain technique also accurately measures the field-like component of spin torque in a wide range of the voltage bias values. 
\end{abstract}
\maketitle

Spin torque (ST) from a direct spin-polarized current \cite{Slonczewski1996,Berger1996,Katine2000,Ralph2008} can excite magnetization auto-oscillations in the free layers of nanoscale spin valves and magnetic tunnel junctions \cite{Kiselev2003,Rippard2004,Nazarov2006, Houssameddine2008, Georges2009, Demidov2010} and thereby generate microwave power \cite{Deac2008,Zeng2012} at a frequency tunable by the current \cite{Kim2008, Kim2008a, Slavin2009a, Silva2010}. Such spin torque oscillator (STO) devices show potential for applications as tunable nanoscale microwave sources and magnetic field sensors for computer hard drives \cite{Braganca2010}. Due to the STOs' nanoscale dimensions, their auto-oscillatory magnetization dynamics are strongly affected by thermal fluctuations \cite{Sankey2005,Muduli2012}, and quantitative understanding of these stochastic dynamics is crucial for the development of devices with desired properties such as narrow generation line width and high frequency agility. 

In this Letter, we demonstrate a method of using time-domain measurements of an STO's voltage oscillations \cite{Krivorotov2005,Nagasawa2011} for quantitative description of the underlying stochastic dynamics. In contrast to frequency domain techniques that probe the  dynamics indirectly via measurements of the STO spectral properties, our method offers a direct look at time evolution of the magnetization vector. We measure the statistical ensembles of the angles at which the magnetization trajectories cross the sample plane, and compare them to predictions made by theories of stochastic magnetization dynamics. Our work demonstrates that the ST-dependent effective energy Fokker-Planck formalism \cite{Apalkov2005a,Bertotti2008} gives an accurate description of the observed ensembles. Based on this effective energy approach, we develop a simple recipe for predicting spectral line widths and line shapes near the generation threshold. This technique also allows us to accurately measure the field-like component of ST (FLT) over a wide range of voltage biases.

\begin{figure}[t!]
\includegraphics[width=\linewidth]{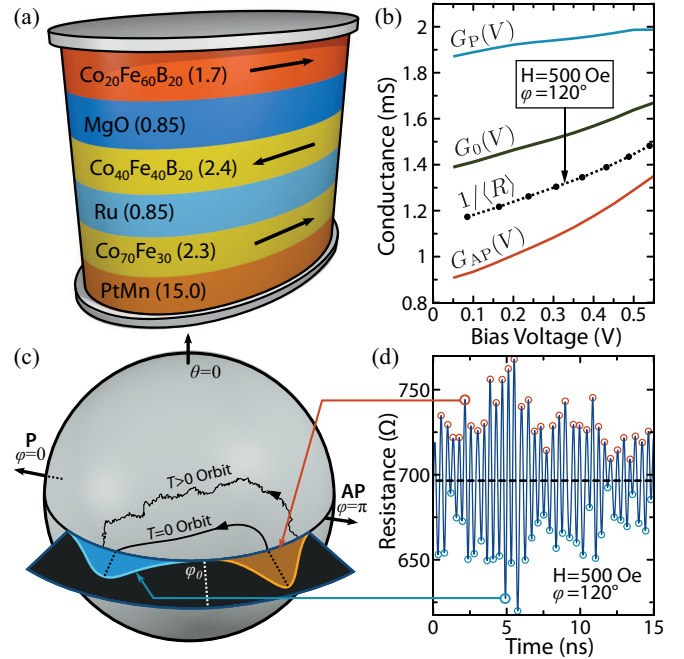}
\caption{\label{fig:stuff}(Color Online) (a) Schematic of the MTJ nanopillar with the layer thicknesses given in nm. (b) Measured bias dependence of the MTJ conductance in the parallel ($G_\text{P}$) and antiparallel ($G_\text{AP}$) states. Also shown are the average conductance $G_0$ and the inverse time-averaged resistance $\langle R \rangle^{-1}$ for the indicated external in-plane field. (c) Examples of trajectories followed by the free layer magnetization on the sphere, and the distributions of the sample plane crossing angles at a non-zero temperature. We use spherical coordinates $\theta$ and $\vp$, where $\theta$ is the polar angle defined with respect to the sample plane normal $\hat{z}$ and $\vp$ is the azimuthal angle defined with respect to the P state direction $\hat{x}$. (d) A time-domain resistance trace, $R(t)$, which maxima and minima are used to reconstruct the sample plane crossing distributions.}
\end{figure}

While the methods discussed here are expected to be general, we focus our present study on STOs based on magnetic tunnel junctions (MTJ) patterned into 150$\times$70 nm$^2$ elliptical nanopillars from a Ta(5)/ PtMn(15)/Co$_{70}$Fe$_{30}$(2.3)/Ru(0.85)/Co$_{40}$Fe$_{40}$B$_{20}$(2.4)/ MgO(0.85)/Co$_{20}$Fe$_{60}$B$_{20}$(1.7)/Ta(5) multilayer (thicknesses in nm). Prior to patterning, the multilayer is annealed for 2 hours at 300$^\circ$C in a 1 Tesla in-plane magnetic field that sets the pinned layer exchange bias direction parallel to the long axis of the nanopillar. The free layer in these structures, pictured in Fig. \ref{fig:stuff}(a), posseses a large perpendicular magnetic anisotropy (PMA) energy $E_\perp = K_1 \sin^2 \theta + K_2 \sin^4 \theta$ that reduces both the critical current $I_c$ and the frequency of the STO auto-oscillations \cite{Ikeda2010, Zeng2012}. Additionally, the free layer exhibits voltage controlled magnetic anisotropy (VCMA) \cite{Maruyama2009,Zhu2012}. The corresponding anisotropy field is $H_{\perp z} = \left[ H_{p0}+\Delta H_{p0}V + H_{p1}\sin^2 \theta \right] \cos \theta$, where $H_{p0}$ is the first order anisotropy field and $\Delta H_{p0}$ is the coefficient of the VCMA field linear in voltage bias.  We include $H_{p1}$, the second order anisotropy field, since it becomes important due to partial cancellation of the out-of-plane shape anisotropy  and the first order PMA \cite{Stamps1997}. The combined effect of PMA and VCMA is, nevertheless, insufficient to overcome the easy-plane magnetic shape anisotropy and the easy axis of the free layer magnetization remains in the sample plane.

In order to extract information on the free layer magnetization trajectories from the STO voltage signal, we find the time-varying total resistance across the MTJ, $R(t)$, which is written as the sum of time-averaged $\langle R \rangle$ and time-dependent $\Delta R(t)$ components. These components are read out, respectively, by a DC voltmeter and a 12 GHz bandwidth 40 GS/s oscilloscope connected to the appropriate ports of a bias-T \cite{supplement}. To ensure that the RF signal far exceeds the noise floor of the oscilloscope (5.2 mV$_\text{rms}$), it is amplified by a 35 dB amplifier with a low noise figure of 1.3 dB . We assume that the angular dependence of the conductance across the MTJ is \cite{Slonczewski2005}
\begin{equation} \label{eqn:conductance} 
G=G_0(V) +\frac{1}{2}\Delta G(V) \m \cdot \p ,
\end{equation}
where $G_0(V)=(G_\text{AP}(V)+G_\text{P}(V))/2$ is the average conductance and $\Delta G(V)=G_\text{P}(V)-G_\text{AP}(V)$ is the full scale conductance change. Here $G_\text{AP}(V)$ and $G_\text{P}(V)$ are, respectively, voltage-dependent conductances in the antiparallel and parallel states of the MTJ, while $\m$ and $\p$ are, respectively, unit vectors in the direction of the fixed and free layers' average magnetizations. We obtain $G_0(V)$ and $\Delta G(V)$, which are plotted in Fig. \ref{fig:stuff}(b), by extracting resistance extrema from hysteresis loops of resistance versus magnetic field taken in the vicinity of the easy-axis as described in the supplemental material \cite{supplement}. The time-dependent component of the resistance is
\begin{equation} \label{eqn:oscRes}
\Delta R(t) = \frac{V(t)}{I} \frac{50\Omega + R_\text{ex} +  \langle R \rangle}{50 \Omega},
\end{equation}
where $I$ is the DC current applied across the device, $R_\text{ex}$ is extrinsic resistance contribution from contacts and sample leads, and $V(t)$ is the voltage signal measured at the $50\Omega$ oscilloscope with the microwave circuit amplification and attenuation factored out. 

\begin{figure}[t]
\includegraphics[width=\linewidth]{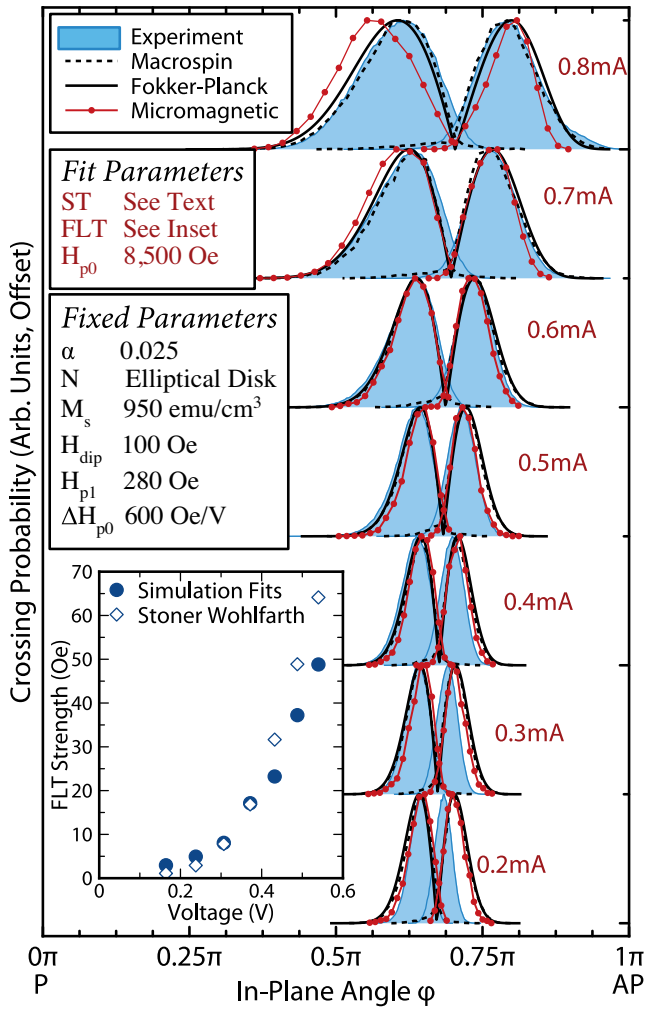}
\caption{\label{fig:dists}(Color Online) Plane-crossing probability distributions $\rho_\text{c}(\vp)$ measured at several values of the bias current for $H=500$ Oe applied at $\vp_H = 120^\circ$. Also shown are distributions calculated from macrospin and micromagnetic simulations as well as by the Fokker-Planck approach described in the text. The macrospin distributions have been shifted towards $\vp = 0$ by $0.03\pi$ as a visual aid. This shift is an artifact of the macrospin approximation \cite{supplement}. The inset shows the magnitude of the FLT extracted (using both macrospin fitting and the Stoner-Wohlfarth model described in the text) from the current-induced shifts of the equilibrium angles.}
\end{figure}

Since the MTJ resistance depends only on the projection of $\m$ onto the polarization vector $\p$, we cannot reconstruct three-dimensional magnetization trajectories (orbits) from the electrical signals. The orbits are, however, symmetric about the sample plane, as seen in Fig. \ref{fig:stuff}(c). Therefore, those points at which $\m$ crosses the equator correspond to extrema in $V(t)$ (and hence $\Delta R(t)$) as pictured in Fig. \ref{fig:stuff}(d). Since the polar angle $\theta = \pi/2$ is known at these crossings, one may determine by means of Eqs. (\ref{eqn:conductance}) and (\ref{eqn:oscRes}) the azimuthal crossing angles
\begin{equation}\label{eqn:mapping}
\vp_i = \cos^{-1} \left[ \frac{2}{\Delta G(V_i)} \left( \frac{1}{R_i} - G_0(V_i) \right) \right]
\end{equation}
Here $R_i$ are the extremal resistance values and $V_i = I R_i$ are the corresponding voltages across the sample. These plane crossing angles are histogrammed separately for crossings at maxima and minima, and are plotted in Fig. \ref{fig:dists} for several values of $I$.
These distributions give new insights into the STO dynamics. As $I$ is raised, the separation between distributions of maxima and minima increase as the ST increases the amplitude of magnetization precession. The individual distributions broaden and become asymmetric due to changes in the ST-dependent effective magnetic energy. The crossing point of the left and right distributions (corresponding to the energy minimum at in-plane angle $\vp_0$) is observed to shift towards the AP state with increasing current. This effect arises exclusively from the FLT, as there are no other voltage dependent fields acting in the plane of the sample.

In order to validate the results of our mapping procedure, we confirm that the  $\vp$ distributions are reproduced by spin torque theory \cite{Slonczewski1996, Berger1996}. The simplest means of generating the expected angular distributions is rote integration of the stochastic Landau-Lifshitz (LL) equation in the macrospin approximation. We make use of a graphics processing unit (GPU) to carry out these calculations for various realizations of the thermal field. The speedup afforded by this method (a factor of at least $10^2$ compared to simulations performed with CPU) allows us to fit the macrospin results directly to the experimental distributions \cite{supplement}. The strengths of the in-plane ST (which pulls $\m$ in the direction $\m \times (\p \times \m)$), FLT (an effective field along $-\p$), and $H_\text{p0} $ are taken to be fitting parameters. We include $H_\text{p0}$ as a global fitting parameter since it exhibits a significant sample-to-sample variation presumably arising from free layer inhomogeneities.  The demagnetization tensor $N$ is assumed to be that of an elliptical disk \cite{Zhu2012,Beleggia2005a}. All additional input parameters, including the Gilbert damping parameter $\alpha$ and the magnitude of the VCMA field, $\Delta H_{p0}$, are taken from independent measurements \cite{Zhu2012}.

\begin{figure}[t]
\includegraphics[width=\linewidth]{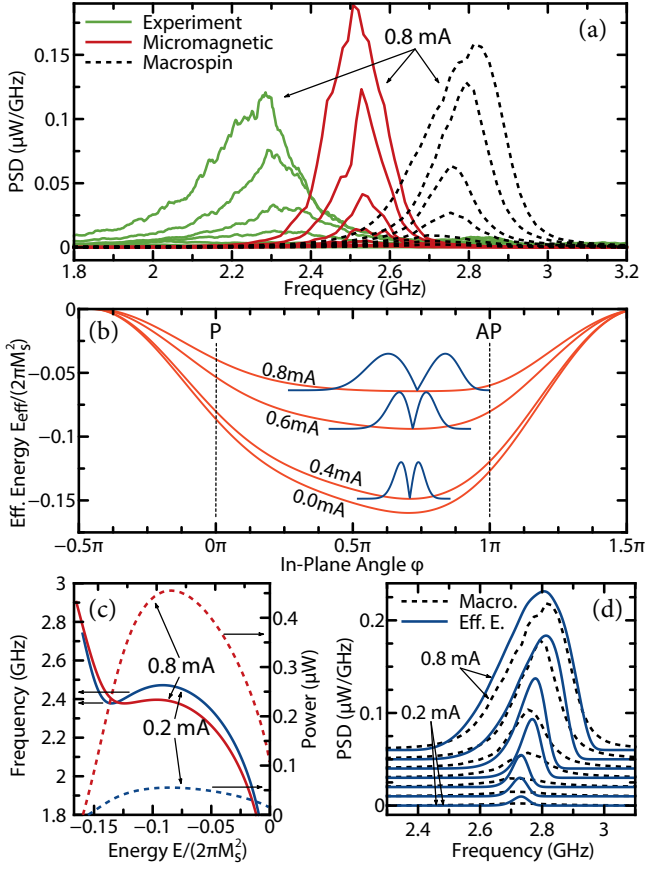}
\caption{\label{fig:spectra}(Color Online) (a) Measured PSDs delivered to a $50 \Omega$ load along with macrospin and micromagnitic simulation results ($I$ in steps of 0.1mA). (b) Effective energies $E_\text{eff}(\vp)$ and corresponding crossing probability distributions $\rho_c(\vp)$. (c) Frequencies (solid lines) and generated electrical powers (dotted lines) of the macrospin conservative orbits. (d) Comparison of PSDs from macrospin simulations and the effective energy approach given by Eq.(\ref{eqn:spectra}).}
\end{figure}

The probability distributions of the plane crossing angles and the generated microwave signal power spectra obtained from these simulations are shown in Figs. \ref{fig:dists} and \ref{fig:spectra}(a). The simulations corroborate our assumption that the current-dependent shift of the distributions is uniquely determined by the FLT, and we are thus able to robustly extract its bias dependence as seen in the inset of Fig. \ref{fig:dists}. In lieu of the fitting method mentioned above, we may calculate the FLT directly from the experimental data in Fig. \ref{fig:dists} by using a simple Stoner-Wohlfarth (SW) model \cite{supplement}. The FLT is thereby
\begin{eqnarray}\label{eqn:SW}
H_\text{flt} &=& -H_x + H_y \cot (\vp_0 + \delta \vp) \nonumber \\
&& + 4 \pi M_s (N_\text{xx} -N_\text{yy}) \cos (\vp_0 + \delta \vp),
\end{eqnarray}
where $H_x$ and $H_y$ are the in-plane components of the external field (including the dipolar contribution), $N_\text{xx}$ and $N_\text{yy}$ are the in-plane components of the demagnetization tensor, $\vp_0$ is the equilibrium angle in the absence of FLT, and $\delta \vp$ is the deviation caused explicitly by FLT. Thus, by recording the deflections of the crossing distributions (at their intersection points), we find that, as shown in the inset of Fig. \ref{fig:dists}, the FLT exhibits an approximately quadratic bias dependence of similar strength to that reported in Ref. \citealp{Wang2011}. In contrast to ST ferromagnetic resonance methods \cite{Tulapurkar2005,Sankey2006}, for which careful subtraction of background signals is needed at non-zero current bias \cite{Wang2011}, Eq. (\ref{eqn:SW}) provides a fast and simple way of measuring FLT strength in a wide range of voltage bias values.

We perform micromagnetic simulations of the STO dynamics in order to confirm that the macrospin approximation adequately describes the system \cite{Donahue1999}. This is indeed the case: the free layer magnetization remains in a macrospin-like state for the studied range of $I$ \cite{supplement}. We show in Figs. \ref{fig:dists} and \ref{fig:spectra}(a) that the micromagnetic results produce power spectral densities (PSDs) and plane crossing angle distributions that are in good qualitative agreement with the macrospin simulations and experimental data. The computational demands of finite-temperature micromagnetic simulations preclude the use of a fitting procedure akin to that employed for macrospin simulations, thus we ran micromagnetic simulations with the parameters identified from the macrospin fits. The exception is the magnitude of the ST polarization efficiency $P$, which was increased to 0.70 from 0.60 in the macrospin case. This discrepancy may stem from both a decrease of the magnetoresistive signal arising from non-uniformities in the free layer magnetization and additional dissipation of the energy supplied by ST drive via micromagnetic degrees of freedom. Despite a small overall shift in the generation frequency ($\approx$ 0.3 GHz), we find that the PSDs given by micromagnetic simulations are in good agreement with the macrospin simulations and the experiment except at the lowest bias currents studied. Local magnetization pinning at defects (discretization artifacts) or nonlinear damping \cite{Tiberkevich2007} may be partially responsible for the discrepancy. 

While providing a direct means of extracting ST parameters from the experiment, simulations of the stochastic LL equation give no insight into the mechanisms leading to the observed asymmetries of the crossing angle distributions and spectral line shapes. We turn to the magnetic Fokker-Planck equation \cite{Brown1963}, which describes the deterministic drift and diffusion of the probability distribution of $\m$ on the unit sphere. Working in the energy coordinate of the system instead of angles $\theta$ and $\varphi$, one can derive \cite{Apalkov2005a,Bertotti2008} a Boltzmann-form probability per unit area of the sphere, $\rho'(E)$, that the system possesses a particular energy, $E$:
\begin{equation} \label{eqn:probE}
\rho'(E,I) = \frac{1}{Z} \exp(- \beta \; \mathcal{V} E_\text{eff}(E,I) ).
\end{equation}
Here $\mathcal{V}$ is the free layer volume, $\beta=1/k_\text{B}T$, and $Z$ is the partition function \cite{supplement}. The ST-dependent effective energy $E_\text{eff}(E,I)$ is constructed by integrating the non-conservative torques acting on the magnetization along the conservative orbit of energy $E$. This approach relies on the assumption that the magnetization mainly evolves along conservative orbits, though it is induced by non-conservative and thermal torques to spread among the orbits on a time scale much longer than that of the oscillation period.

For comparison with our measurements, we are interested not in $\rho'(E)$, but rather in the probability $\rho_\text{c}(\vp)$ that $\m$ crosses through the plane of the sample at angle $\vp$. This quantity is given by
 \begin{equation}\label{eqn:probCross}
  \rho_\text{c}(\vp) = \frac{2 \pi \gamma M_s}{Z} \frac{\rho'(E(\vp),I)}{\omega(E(\vp))} \left| \deriv{E(\vp)}{\vp} \right|,
 \end{equation}
where $\omega(E)$ is the angular frequency and $E(\vp)$ is the in-plane cross-section of the conservative energy surface \cite{supplement}. The distributions $\rho_\text{c}(\vp)$ and energy surfaces $E_\text{eff}(E)$ from which they are derived are shown in Fig. \ref{fig:spectra}(b). The Fokker-Planck approach necessarily predicts zero probability at the minimum of $E(\varphi)$ due to the vanishing density of states $ \left| \dd E(\vp) / \dd \vp \right|$ appearing in Eq. (\ref{eqn:probCross}). The crossing distributions are also plotted in Fig. \ref{fig:dists}, wherein we see excellent agreement with the macrospin results over the entire range of currents.  We note that Eq. (\ref{eqn:probE}) and Eq. (\ref{eqn:probCross}) can be easily inverted to reconstruct the effective energy surface $E_\text{eff}(E,I)$ from the measured plane crossing probability distributions $\rho_\text{c}(\vp)$.

We can now identify the cause of asymmetry and broadening of $\rho_c(\vp)$ observed at large values of $I$. While $E_\text{eff}(\vp)$ is approximately quadratic near the bottom of the well, it quickly crosses into a nearly linear regime with increasing $\varphi$. The number of available orbits between $E$ and $E+\dd E$ becomes large in this region of $E_\text{eff}(\vp)$ and $\m$ spends proportionally more time on these large amplitude trajectories. This causes the tail of $ \rho_\text{c}(\vp)$ to elongate. As the slope of this linear region decreases in response to increasing $I$, the distributions widen and distort commensurately.

For currents near $I_c$, flatness at the bottom of the effective energy well $E_\text{eff}(\vp)$ suggests a simple method for evaluating the STO's PSD. The relaxation of $\m$ towards its equilibrium orbit is driven by deterministic torques proportional to the slope of $E_\text{eff}(\vp)$. Thus, near the bottom of $E_\text{eff}(\vp)$, these restoring torques become small and the time evolution of the oscillation amplitude becomes dominated by thermal diffusion. If the time scale for thermal diffusion of the amplitude is long compared to the period of oscillations, the PSD can be approximated by a superposition of the power generated by each orbit weighted by the probability of finding $\m$ on this orbit given by the simple Boltzmann-like expression of Eq. (\ref{eqn:probE}). We proceed with this simple method for determining the PSD, and thereby verify that this diffusion-dominated limit is applicable for our system. We first compute the frequencies $\omega(E)$ and average electrical powers $P(E)$ (delivered to a $50\Omega$ load) corresponding to each conservative orbit, both of which are plotted in Fig. \ref{fig:spectra}(c). For a single-valued function $E(\omega)$, the PSD is given by the electrical power of individual orbit $P(E)$ weighted by the probability of finding the system at this orbit:
\begin{equation}\label{eqn:spectra}
S(\omega) = \frac{2 \pi \gamma M_s \rho'(E(\omega),I) P(E(\omega))}{\omega}.
\end{equation}
For a multi-valued function $E(\omega)$, such as that shown in Fig. \ref{fig:spectra}(c), a sum over all branches of $E(\omega)$ should be added to the right hand side of Eq. (\ref{eqn:spectra}). As shown in Fig. \ref{fig:spectra}(d), the PSDs calculated with Eq.(\ref{eqn:spectra}) possess powers, line widths, and line shape asymmetries that are in excellent agreement with the macrospin simulations and experimental PSDs for $I$ close to $I_c$. Far from $I_c$, the agreement between PSD given by the macrospin simulations and by Eq.(\ref{eqn:spectra}) decreases because rapid amplitude relaxation and an associated increase in phase noise (arising from nonlinear coupling between oscillation amplitude and phase \cite{Slavin2009a}) belie the simplicity of our ensemble average. 

With Eq. (\ref{eqn:spectra}) we can now see that the PSD broadening and asymmetry near $I_c$ has the same origin as the asymmetry of the crossing angle distributions $\rho_c(\vp)$. The occupation of larger amplitude orbits increases substantially as $E_\text{eff}(\vp)$ develops a large linear region. As seen in Fig. \ref{fig:spectra}, these orbits posses lower frequencies due to the frequency redshift with increasing amplitude. Thus the PSDs are seen to spread asymmetrically towards lower frequencies. In principle, non-monotonicities in $\omega(E)$ such as those seen in Fig. \ref{fig:spectra}(b) can create, by means of an increased density of states, an accumulation of power near local extrema of $\omega(E)$. Closely spaced features of this sort could even give the illusion of multiple modes. Since our samples are prone to dielectric breakdown at higher current densities, we are unable to experimentally access this regime.

In conclusion, time domain measurements of STO voltage allow us to rapidly map statistical ensembles of STO magnetization trajectories and thereby determine ST-dependent Fokker-Planck effective energy of the STO. We use the ST-dependent effective energy approach to derive a simple expression for the STO power spectral density valid at $I\approx I_c$, which is in excellent agreement with our experimental data. Our technique also provides a simple and accurate measurement of the field-like component of ST across a wide range of voltage bias values. 

We would like to thank M. W. Keller for initial discussions on the feasibility of examining real-space magnetization dynamics. We also thank A. Slavin, V. Tiberkevich, P. Visscher, P. Braganca, and B. Gurney for useful discussions. This work was supported by DARPA Grant No. HR0011-10-C-0153 and by NSF Grants No. DMR-1210850 and No. ECCS-1002358. 

%\bibliography{cleaned-abbrev}

\end{document}

\title{Time Domain Mapping of Spin Torque Oscillator Effective Energy: \\
Supplemental Material}
\author{Graham E. Rowlands}
\affiliation{Department of Physics and Astronomy, University of California, Irvine, California 92697, USA}
\author{Jordan A. Katine}
\affiliation{Hitachi Global Storage Technologies, 3403 Yerba Buena Road, San Jose, CA 95135}
\author{Juergen Langer}
\affiliation{Singulus Technologies, 63796 Kahl am Main, Germany}
\author{Jian Zhu}
\affiliation{Department of Physics and Astronomy, University of California, Irvine, California 92697, USA}
\author{Ilya N. Krivorotov}
\affiliation{Department of Physics and Astronomy, University of California, Irvine, California 92697, USA}\date{\today}
\maketitle

\tableofcontents

\section{Reconstructing the Real-Space Dynamics}
\begin{figure*}[htb]
\includegraphics[width=\linewidth]{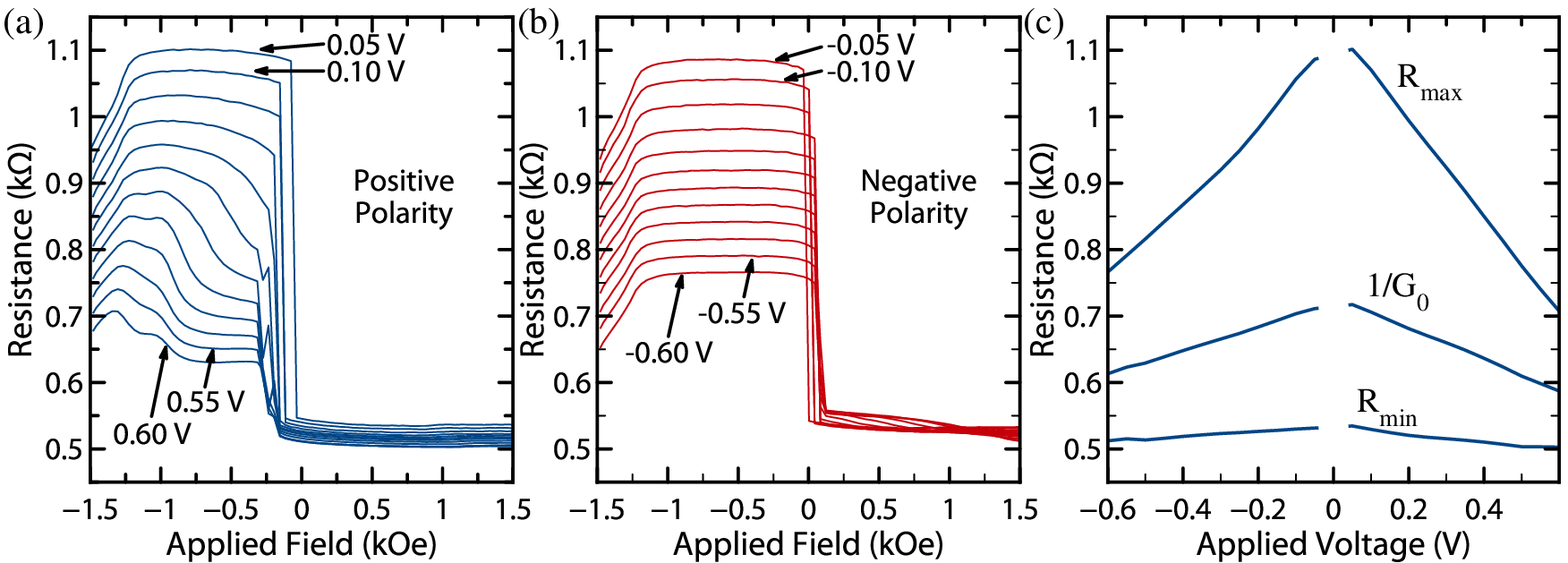}
\caption{\label{fig:extraction}(Color Online) Resistance as a function of field at various values of (a) positive and (b) negative bias voltage. The increment between adjacent traces is 50mV.  (c) The corresponding $R_\text{AP}(V)$, $R_\text{P}(V)$, and $1/G_0(V)$ as extracted from (a) and (b). The extrinsic resistance contribution $R_\text{ext}$ is included in (a--c), though it is subtracted for the actual analysis. }
\end{figure*}

Accurate determination of the spatial orientation of the free layer magnetization $\mm$ is predicated upon careful characterization of a sample's voltage-dependent tunneling magnetoresistance (TMR). To this end, the maximum and minimum observed resistances $R_\text{AP}(V)$ and $R_\text{P}(V)$ are extracted as a function of bias from hysteresis loops taken in the vicinity of the sample's easy axis. By measuring within a $\pm 5^\circ$ range of angles we are able to account for potential field misalignments. Representative hysteresis loops are shown for both voltage polarities in Figs. \ref{fig:extraction}(a) and \ref{fig:extraction}(b), wherein the influence of magnetization dynamics on the average resistance of the sample is readily visible (positive current pulls the magnetization towards the parallel state in this geometry). We have confirmed that the powers of these magnetization oscillations are negligible at high fields (approaching 1.5 kOe), and therefore conclude that the opening angles of the trajectories are too small to interfere with measurements of $R_\text{P}(V)$ and $R_\text{AP}(V)$.

The bounding resistance values, as well as the inverse of the average conductance $G_0^{-1}$ (corresponding to an angle $\vp=\pi/2$ between $\m$ and $\p$) are plotted in Fig. \ref{fig:extraction}(c) as a function of bias. These traces provide all of the necessary information for mapping the sample's resistance state back to its orientation by means of Eq. 3 of the main text.

\begin{figure}[htb]
\includegraphics[width=\linewidth]{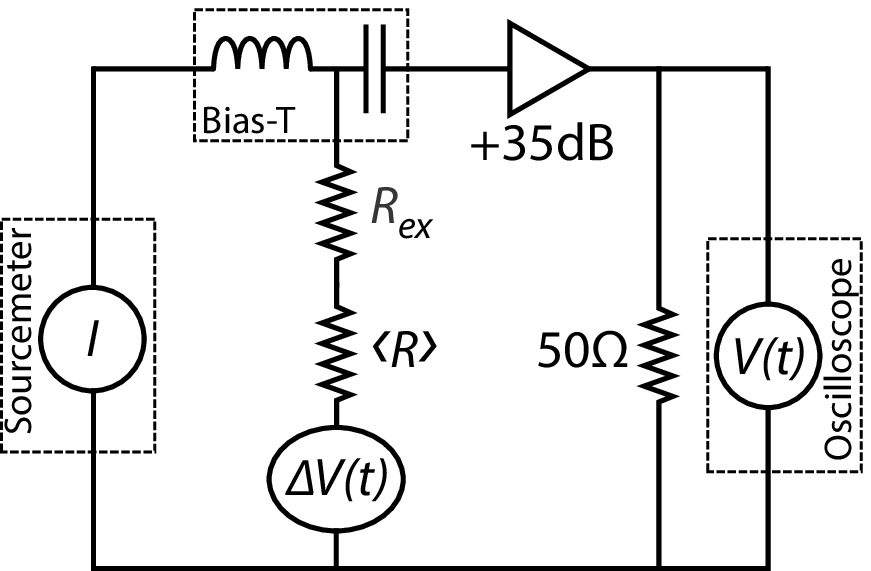}
\caption{\label{fig:circuit}(Color Online) The microwave circuit used for time domain analysis. The same instrument is used to supply current $I$ and measure the time averaged resistance across the DC branch of the circuit. The amplified microwave voltage generated by the sample, $\Delta V(t)$, results in the voltage $V(t)$ across the characteristic 50$\Omega$ real time oscilloscope.}
\end{figure}

Obtaining the time-dependent sample resistance still requires some care. We must determine the voltage oscillations across the sample $\Delta V(t)$ corresponding to the signal $V(t)$ measured at the oscilloscope. This requires characterization of our microwave circuit as detailed in Fig. \ref{fig:circuit}. After accounting for active amplification and passive circuit attenuation, the desired signal is given by
\begin{equation}
\Delta V(t) = V(t) \frac{50\Omega + R_\text{ex} + \langle R \rangle}{50\Omega},
\end{equation}
where $R_\text{ex}=21\Omega$ is the resistance contribution from sources extrinsic to the sample itself (from contacts and probes), and $\langle R \rangle$ is the time averaged MTJ resistance (less the extrinsic contribution) that we measure using the sourcemeter that supplies direct current to the device. 
The voltage oscillations measured by the oscilloscope may then be mapped back to the angular separation of the free layer and polarizer. As described in the main text, we can only pinpoint the location of the magnetization as it crosses the plane of the sample, which corresponds to extrema in $R(t)$. It remains, then, to develop a robust peak-finding procedure in order to correctly identify the extrema from $R(t)$. 

\begin{figure}[htb]
\includegraphics[width=\linewidth]{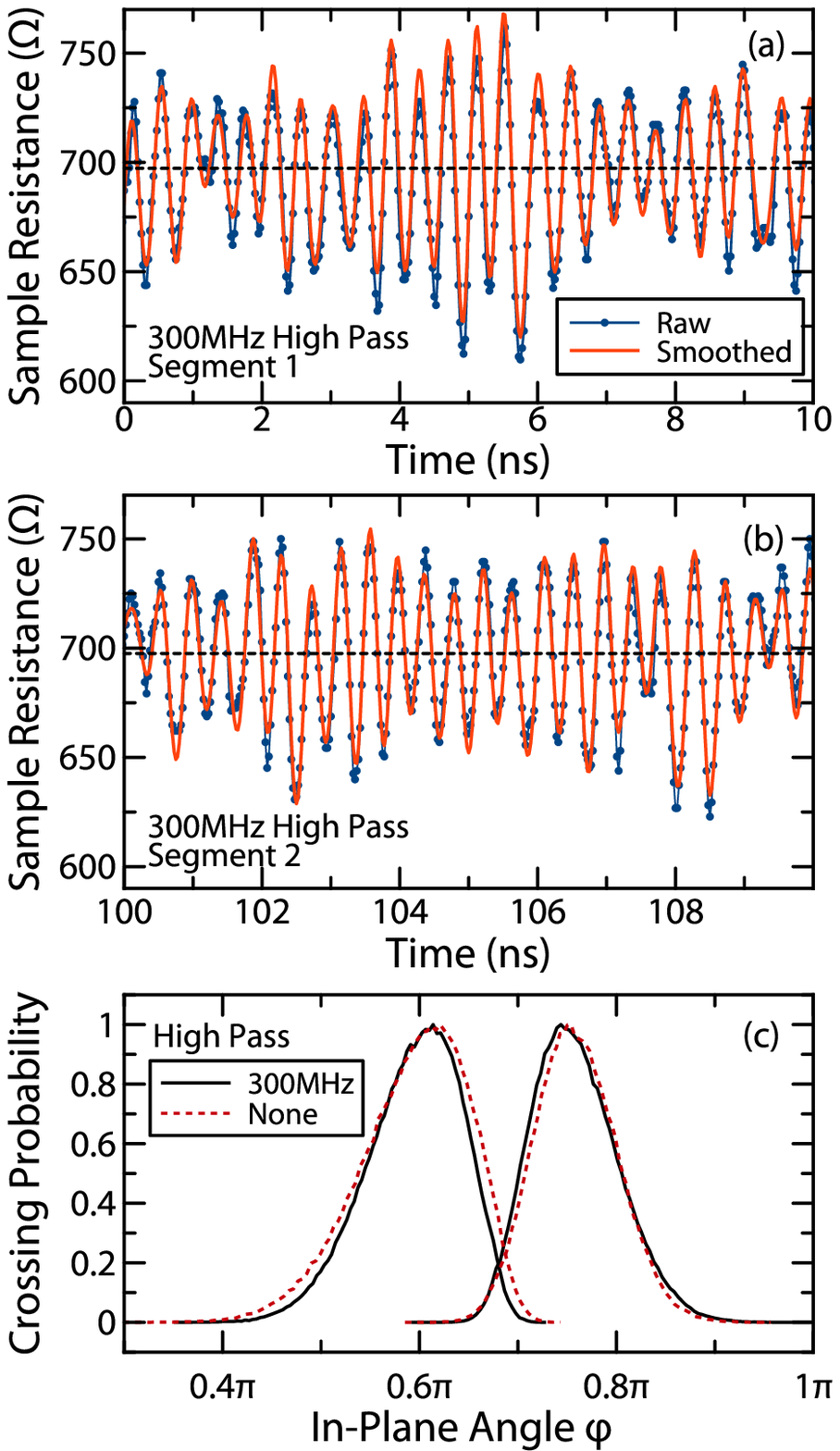}
\caption{\label{fig:smoothing}(Color Online) Two representative segments (a,b) of raw and band-pass filtered (``Fourier smoothed'') time-domain data. The horizontal dotted line represents the time-averaged resistance $\langle R \rangle$. Local extrema are seen to be effectively removed by low-pass filtering. The effects of high-pass filtering are shown in (c) where the reconstructed crossing distributions are shown with and without such a filter. All data are shown for $I=0.7$mA, $H=500$Oe, and $\vp=120^\circ$.}
\end{figure}

The primary culprits of reconstruction errors are erroneous extrema originating from any source of high-frequency low-amplitude noise, such as thermal fluctuations of the magnetization orientation or an extrinsic source such as electronics noise. Such anomalies are easily eliminated by the use of a Fourier smoothing procedure that eliminates high frequency features in the voltage signal. We low-pass filter all frequency components greater than 1.5 times the fundamental frequency of the oscillator, beyond which the experimentally measured power spectral density is nearly zero. We apply a magnetic field of sufficient strength and appropriate direction to intentionally constrain the ensemble of orbits away from the easy axis. Otherwise a second harmonic would be present in the voltage signal, leading to a more complicated crossing angle measurement procedure. Also, since low-frequency noise can smear out the reconstructed angular distributions, we only admit frequencies above 300 MHz, below which point the noise figure of our amplifier rapidly increases from its base value of 1.3dB. We do not observe any large low-frequency features in the experimentally obtained spectra of the STO, and therefore do not expect that this high-pass filtering will impact the accuracy of the method. Examples of both raw and band-pass filtered time-domain data are shown in Fig. \ref{fig:smoothing} (a, b) along with an example of how the low-pass filtering affects the crossing angle distributions (Fig. \ref{fig:smoothing}(c)).

A severe micromagnetic curvature of the free layer would preclude use of our method, since the projection $\m \cdot \p$ would no longer provide a complete representation of the free layer's state. Later on in this supplement we present the results of micromagnetic simulations that suggest there is minimal distortion of the magnetization profile from the macrospin state.

A broadening of the reconstructed peaks will inevitably result both from small micromagnetic distortions and from the noise at the oscilloscope. We therefore expect that the angular distributions that we obtain are actually convolutions of the intrinsic distributions with an approximately Gaussian kernel. Deconvolution is unavailing in this particular context, so we stress that the distributions we plot are slightly smeared versions of the true ones.  This effect is most pronounced for small applied currents since the relative contribution of the electronics noise ($\approx$ 5 mV$_\text{rms}$) is largest in that case. At most we expect a 3--4$^\circ$ blurring of the distributions, which does not alter any of their qualitative features. 

\section{Macrospin Simulations}

Macrospin simulations allow us to make quick comparisons of our experimental results to theoretical expectations. We adopt the stochastic Landau-Lifshitz (LL) form of the magnetization dynamics \cite{Bertotti2008},
\begin{equation}\label{eqn:ll}
\dd \mm =  \left[   \mathbf{v}(\mm,t)  -\nu^2 \mm \right]\dt  - \mm \times \left(     \nu \dW + \alpha \nu \mm \times \dW    \right),
\end{equation}
interpreted in the It\={o} sense \cite{Gardiner2009}. The deterministic part of the dynamics is described by
\begin{equation}\label{eqn:ll2}
\mathbf{v}(\mm,t) = - \mm \times \left( \heff + \alpha  \mm \times \left( \heff + \beta_{\text{st}} \pp / \alpha \right) \right),
\end{equation}
where $\alpha$ is the Gilbert damping parameter, $\nu$ is the magnitude of thermal fluctuations as given by the fluctuation-dissipation theorem, and $\text{d}\mathbf{W}$ is the isotropic Wiener process that is the generator of Brownian motion. All fields and magnetic moments are normalized by the saturation magnetization $M_s$. In Eqs. (\ref{eqn:ll}) and (\ref{eqn:ll2}) only, the units of time are are multiplied by  $\gamma M_s$ (where $\gamma$ is the gyromagnetic ratio) to yield a dimensionless quantity. The spin-torque coefficient $\beta_{\text{st}}$ depends on the mutual angle of $\mm$ and $\pp$ and is given by \cite{Sun2008}
\begin{equation}
	\beta_{\text{st}} = \frac{a_{\text{st}}  I \hbar}{2 e M_s^2\mathcal{V}}\left( \frac{P}{1+P^2 \mm \cdot \pp} \right), 
\end{equation}
where $P$ is the polarization efficiency, $I$ is the current, and $\vol$ is the free layer's volume. The dimensionless factor $a_{\text{st}}$ gives the strength of the in-plane spin torque, and is one of the fitting parameters used when comparing the simulation results to experimental data. The field-like torque
\begin{equation}\label{eqn:flt}
	\mathbf{h}_{\text{flt}} = \frac{a_{\text{flt}} I \hbar}{2 e M_s^2 \mathcal{V}}\left( \frac{P}{1+P^2 \mm \cdot \pp} \right) \pp
\end{equation}
is taken to be a constituent of the effective field $\heff$. Here, $a_{\text{flt}}$, represents the dimensionless strength of the field-like torque. We integrate Eq. (\ref{eqn:ll}) over many realizations of the Wiener process simultaneously on a GPU using an Euler algorithm with appropriately small time increment $\Delta t = 30.\times10^{-15}$ s. The probability distributions calculated from these newly alacritous simulation results can be fed directly into a least-squares optimization algorithm. The simulation results are interpolated onto the same set of $\vp$ points as the experimental distributions, and the residuals between the two are minimized with respect to $a_{\text{stt}}$, $a_{\text{flt}}$, and the first-order perpendicular magnetic anisotropy field $H_{p0}$. The plots of the FLT strength in the main text is given, instead, in units of field by utilizing the whole prefactor of Eq. (\ref{eqn:flt}). 

The spectral properties of these dynamics are quantities of great interest. In order to compare to experimentally observed power spectral densities, we must calculate the actual voltage signal that the simulated device would deliver, \emph{ceteris parabis}, to a $50\Omega$ load in the experimental circuit. This procedure is somewhat complicated by the nature of the measurement: in the presence of a constant current $I$ across the device, changes in the orientation of the free layer alter the voltage across the junction, which in turn changes the observed resistance through the bias-dependence of TMR. We must therefore derive an expression for the voltage as a function of both $I$ and $\m$. Starting from the expression for the conductance as a function of angle and voltage,
\begin{equation} \label{eqn:conductance} 
G(V,\theta)=\frac{I}{V} = G_0(V) +\frac{1}{2}\Delta G(V) \mm \cdot \pp,
\end{equation}
and approximating the functions $G_0(V)$ and $G(V)$ as
\begin{subequations}
\begin{align}
G_0(V) &= g_{0} +  g_{1}V  + g_2 V^2\\
\Delta G(V) &= h_{0} +  h_{1}V + h_2 V^2 ,
\end{align}
\end{subequations}
we solve this system of equations for $V$ and expand in powers of $\cos(\theta)$ and $I$ to whichever order is necessary for a reasonable facsimile to the full solution. Using the final approximation of $V(I,\m)$ we calculate the voltage signal at $N$ individual points $V_i$ separated by time $\Delta t$. These points are then multiplied by a Hann window in order to mitigate spectral artifacts caused by the finite signal lengths. The RMS power delivered to a $50\Omega$ load at spectral component $k$ is then
\begin{equation}
S_k = \frac{1}{2} \left| \frac{ 2 \mathcal{F}_k[V_i \cdot w_i] }{N} \right| ^2 \left( \frac{50\Omega}{50\Omega + R_\text{ex} + \langle R \rangle } \right) ^2 \frac{1}{50\Omega}
\end{equation}
where $\mathcal{F}[x_i]$ represents the (un-normalized, single-sided) discrete Fourier transform, $w_i$ are the sampled points of the Hann windowing function, and the factor of $2$ in the numerator accounts for the power lost in the first harmonic from the use of the windowing function. The power spectra are calculated separately for each realization of the thermal field, and are then averaged to obtain the final spectrum. 

\section{Extracting the Field-Like Torque}

Extraction the field-like torque from crossing distributions involves many fewer complications than deriving the same results from ST ferromagnetic resonance (STFMR) methods. We assume a Stoner-Wohlfarth (SW) energy density of the form
\begin{eqnarray}\label{eqn:SW}
E(\vp) &=& 2 \pi M_s^2 (N_{\text{xx}} - N_{\text{yy}}) \cos^2 \vp \nonumber \\
&& - (H_x + H_\text{flt}) M_s \cos \vp - H_y M_s \sin \vp,
\end{eqnarray}
where $H_x$ and $H_y$ are the in-plane components of the external field (including the dipolar contribution $H_\text{dip}$), $N_\text{xx}$ and $N_\text{yy}$ are the in-plane components of the demagnetizing tensor, and $H_\text{flt}$ is the FLT contribution we seek to identify. We find the equilibrium azimuthal angle $\vp_0$ that minimizes Eq. (\ref{eqn:SW}) with $H_\text{flt}=0$, then we express the magnitude of the FLT 
\begin{eqnarray}
H_\text{flt} &=& -H_x + H_y \cot (\vp_0 + \delta \vp) \nonumber \\
&& + 4 \pi M_s (N_\text{xx} -N_\text{yy}) \cos (\vp_0 + \delta \vp)
\end{eqnarray}
in terms of the FLT-induced deflection from equilibrium $\delta \vp$. By comparing the displacements, rather than absolute angles, we can partially account for errors (discussed below) induced a slight micromagnetic curvature of the system. It would be natural to compare the results furnished by this method to those from STFMR, though we were unable to do so for our sample given the presence of large backgrounds in our finite-bias resonance spectra (a typical complication of such measurements). 

\section{Micromagnetic Simulations}

A more realistic representation of the experimental system is obtained in the micromagnetic approximation. To this end we modify the OOMMF finite-difference micromagnetic framework to perform finite temperature simulations of the LL equation including spin-torque in a form appropriate for MTJs \cite{Donahue1999}. The computational burden of these simulations is great: Gaussian random deviates must now be generated for each site in the discretized system, and the time step of the simulations must be drastically reduced (from the zero temperature case) in order that the timescale of thermal fluctuation remains much smaller than the timescale of the magnetization dynamics.

\begin{figure}[h]
\includegraphics[width=1.0\linewidth]{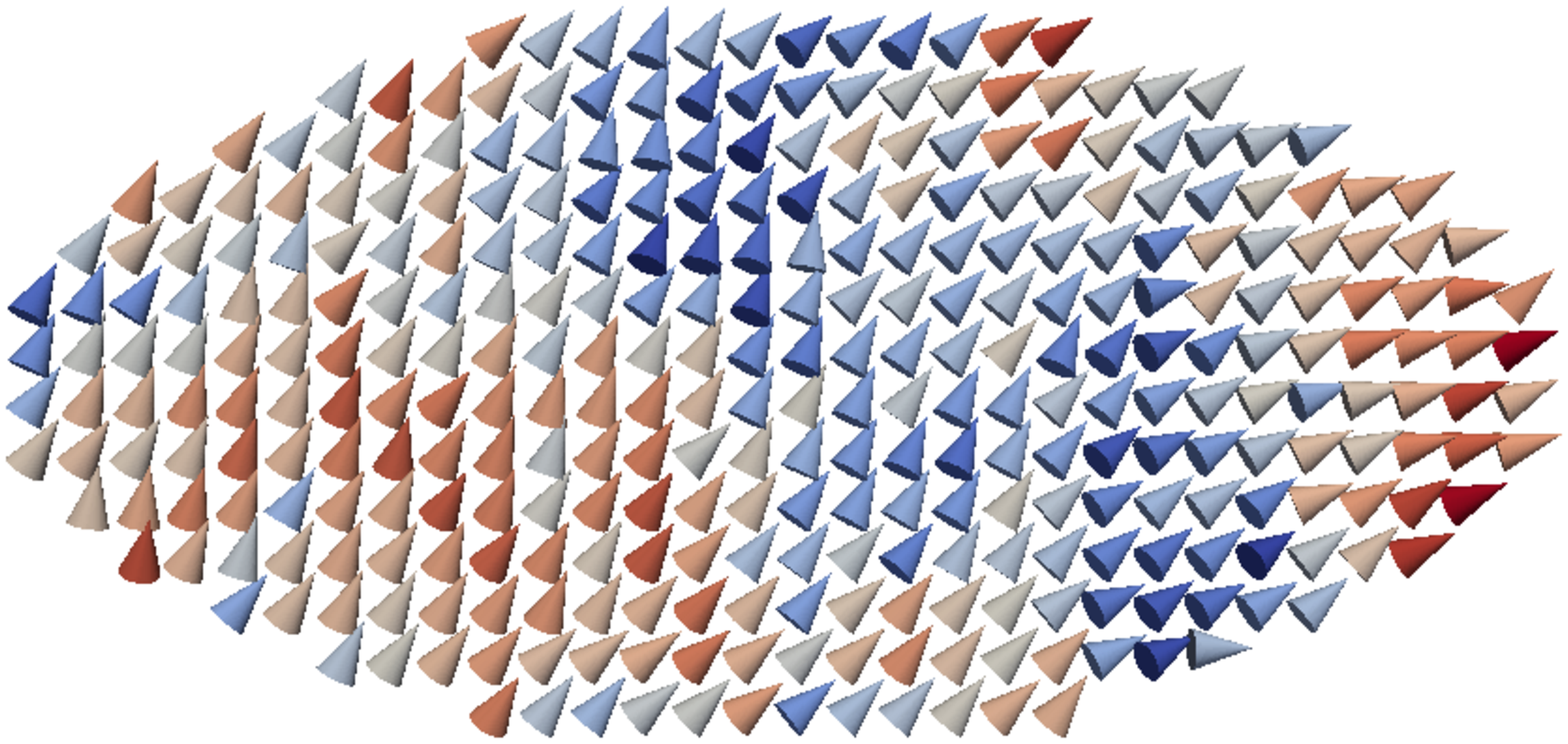}
\caption{\label{fig:Micromagnetic}(Color Online) Snapshot of the micromagnetic configuration of the sample at $I=0.8$mA (the largest current value used in this study), wherein color encodes the out-of-plane component of magnetization.}
\end{figure}

For these simulations we set the exchange stiffness $A$ to $2.0\times10^{-6}$ emu$/$cm, while the remaining simulation parameters are identical to those used in the macrospin approximation. We deliberately omit a realistic magnetostatic field from the polarizer in favor of a simple uniform field for the sake of a more straightforward comparison to the macrospin results. The aim of this work is not to reproduce the exact local environment of the experimental system, but rather to gauge the agreement of these various numerical and analytical methods.

During the course of these simulations the average magnitude of the magnetization unit vector (as defined by the norm of the vector average over cells) spends approximately 50\% of its time above 0.95 and approximately 90\% of its  time above 0.90. This is a good indication that the micromagnetic curvature of the system has a minimal impact on our reconstruction process. We show in Fig. \ref{fig:Micromagnetic}, for reference, a high-curvature snapshot of the magnetization along a large amplitude trajectory. 

The micromagnetic simulations resolve an apparent inconsistency between the experiment results and the macrospin simulations: the equilibrium angles predicted within the macrospin approximation (i.e. by the Stoner-Wolfarth model) do not match those observed in experiment; they differ by a factor of $0.03 \pi$. This discrepancy is due to a slight micromagnetic curvature of the system, though we emphasize that this curvature found in static equilibrium is even smaller than dynamic curvature observed during large-amplitude motion of the magnetization pictured in Fig. \ref{fig:Micromagnetic}.

\section{The Effective Energy Approach}

\begin{figure*}[ht]
\includegraphics[width=1.0\linewidth]{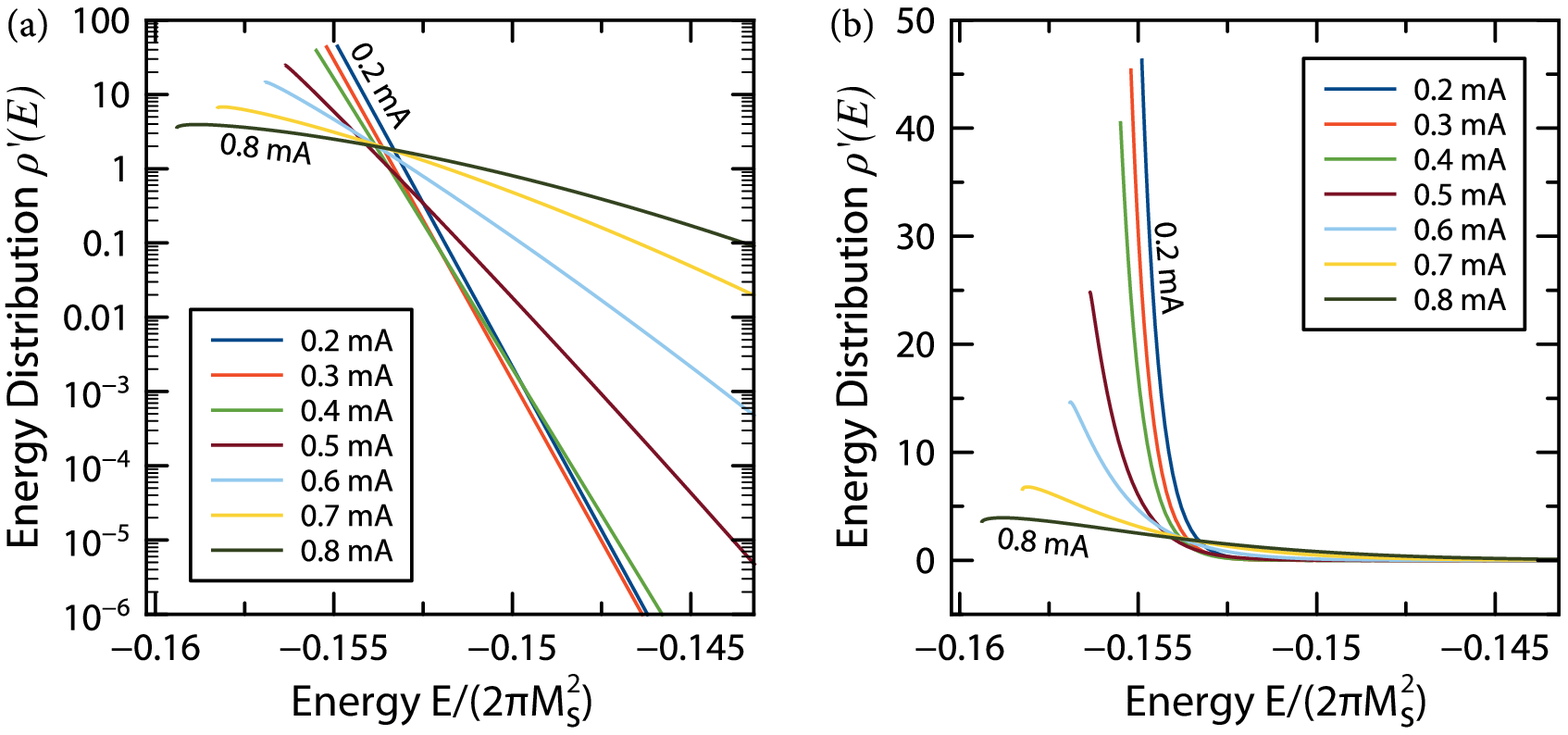}
\caption{\label{fig:dists}(Color Online) (a) Logarithmic and (b) linear scale plots of the energy distributions according to Eq. (\ref{eqn:energyDist}). The distributions are seen to diverge slowly from the Boltzmann result at $I=0$. The minimum available energies at successive currents are observed to shift. This is due to modifications of the energy surface by FLT and voltage induced anisotropy, which together decrease the lowest attainable energy value with increasing $I$. One can also observe small peaks in the high-bias distributions just above the energy minima. These are indications that the system has reached the critical current and favors an orbit of non-zero radius corresponding to an energy above that at the bottom of the well. }
\end{figure*}
The aforementioned simulation techniques do little to further our understanding of the system beyond what is already apparent from measurements. We seek to understand, primarily, the origin of asymmetry and broadening in both the $\vp$ distributions and spectral lines. An analytical approach is therefore sought. Since we are directly measuring probability distributions, it behooves us to leave behind the stochastic differential equation Eq. (\ref{eqn:ll}) in favor of the equivalent Fokker-Planck equation. Following the original derivation of the magnetic Fokker-Planck equation \cite{Brown1963}, it has since been recast in the effective energy coordinate of the system \cite{Apalkov2005a,Bertotti2008}. The central idea of this approach is to assume that non-equilibrium and thermal torques cause the system to drift slowly between conservative orbits of the system (which are by definition calculated in the absence of such drives). For a magnetic system with a single potential well, the effective energy associated with a conservative orbit of energy density $E$ is 
\begin{equation}
E_\text{eff}(E) - E_\text{eff}(E_0) = (E- E_0) - I \int_{E_0}^{E}  \eta(E') \text{d} E' ,
\end{equation}
where $E_0$ is an arbitrary reference energy which we choose to be the maximum of the in-plane energy $E(\vp)$. The factor $I \eta(E)$ is the ratio of the works done by ST and damping over the conservative orbit of energy $E$. The current $I$ has been factored out in order to show the explicit current dependence. The functional form of $\eta(E)$ can be found from the constituent torques of Eq. (\ref{eqn:ll}). Aside from energy contributions of the demagnetizing and external fields, we include those due to perpendicular anisotropy and field-like torque (FLT), which we treat as an effective field contribution. The energy density from perpendicular anisotropy is given by
\begin{equation}
E_{\text{an}}  = - K_1 \sin^2\theta -K_2 \sin^4 \theta ,
\end{equation}
where $K_1$ and $K_2$ are the first and second order anisotropy energies:
\begin{eqnarray}
K_1 &=& \frac{1}{2}(H_{p0} + \Delta H_{p0}) M_s  \\
K_2 &=& \frac{1}{4} H_{p1} M_s.
\end{eqnarray}
In addition to the first and second order anisotropy fields, $H_{p0} $ and $H_{p1}$, we include the voltage-controlled magnetic anisotropy (VCMA) contribution $\Delta H_{p0}$ which we take to be 600 Oe/V based on experiments conducted on similar samples \citep{Zhu2012}. The field-like torque's energy contribution is, meanwhile, given by \cite{Bertotti2008}
\begin{equation}
E_{\text{flt}}  =- \frac{a_{\text{flt}} I \hbar}{2 e P \vol}\ln \left( 1+P^2 \mm \cdot \pp \right).
\end{equation}
When including either of these voltage and current-dependent terms we must recalculate the set of conservative orbits at each current step since they modify the effective field. We employ our aforementioned GPU-based macrospin simulation code to simultaneously calculate the entire set of conservative orbits. The probability per unit area of the unit sphere of finding the magnetization at energy $E$ is given in the convenient Boltzmann form:
\begin{equation}\label{eqn:energyDist}
%\rho'(E) = \frac{1}{Z} \exp \left( - \beta \; \vol \left[ E_\text{eff}(E) - E_\text{eff}(E_0) \right]  \right),
\rho'(E,I) = \frac{1}{Z} \exp(- \beta \; \mathcal{V} E_\text{eff}(E,I) ).
\end{equation}
where $\beta $ is the inverse of the thermal energy $(k_{B} T)^{-1}$, and the current has been added as an argument of the effective energy surface since it must be recalculated for every choice of $I$. The factor
\begin{equation}\label{eqn:partition}
Z = \gamma M_s \int  \tau(E') \rho'(E',I) \dd E'
\end{equation}
is a ``non-equilibrium partition function" resulting from an integral over all energy states (see Eq. 10.203 and related discussion of Ref. \citealp{Bertotti2008}). Here, $\tau(E)$ are the periods corresponding to conservative orbits of energy $E$. The $\rho'(E)$ distributions are shown for several values of the current in Fig. \ref{fig:dists}, where one can readily observe the transition from Boltzmann to non-Boltzmann behavior with increasing $I$. We are interested in the probability that the magnetization crosses through the plane of the sample at angle $\vp$. This quantity is simply related to $\rho'(E)$, and is found through the ``conservation of probability'' to be
 \begin{equation}\label{eqn:probCrossProto}
 \rho_\text{c}(\vp) = \frac{1}{Z} \rho'(E(\vp),I)   A(E(\vp))    \left| \deriv{E(\vp)}{\vp} \right|.
 \end{equation}
In this expression, the density of states accounts for the number of orbits between $\vp$ and $\text{d}\vp$, while $E(\vp)$ denotes the in-plane cross-section of the energy surface with $\theta = \pi/2$. The crossing probability $\rho_\text{c}(\vp)$ always goes to zero at the center of the well since the density of states $\left| \dd E(\vp) / \dd \vp \right|$ vanishes at that point. The factor $A(E) \dd E = \gamma M_s \tau(E) \dd E$ gives the surface area of the unit sphere enclosed by the band of orbits between $E$ and $E+\dd E$, and is proportional to the period of the orbits $\tau(E)$ in that vicinity (as shown in the supplemental material of Ref. \citealp{Apalkov2005a}). The area factor, a minor correction for our system, is necessary since $\rho'(E)$ is specified per unit area of the unit sphere. We may rewrite the crossing probability in a more intuitive form as
\begin{equation}\label{eqn:probCross}
 \rho_\text{c}(\vp) = \frac{2 \pi \gamma M_s}{Z} \frac{\rho'(E(\vp),I)}{\omega(E(\vp))} \left| \deriv{E(\vp)}{\vp} \right|,
\end{equation}
where $\omega(E)$ is the angular frequency. Thus, we need only record $\omega(E)$ and $P(E)$ (the power dependence) in order to furnish predictions for the spectral line shapes of the STO. This is an important result, as such quantities are readily obtained from simulations.

In the approximation of this model (as described in the main text), the line shape is given by the sum of the powers of individual orbits weighted by their relative occupation probabilities:
\begin{equation}\label{eqn:spectra1}
S(\omega) = \rho'(E(\omega),I) A(E(\omega)) P(E(\omega)).
\end{equation}
We can make a transformation from $A(E)$ to $2\pi \gamma M_s / \omega(E)$ to yield the expression in the main text:
\begin{equation}\label{eqn:spectra2}
S(\omega) = \frac{2 \pi \gamma M_s \rho'(E(\omega),I) P(E(\omega))}{\omega}.
\end{equation}
The application of this method to the experimental system described in the main text has proved very successful near threshold. It remains that we should compare this method to a well-established formalism that is widely used to describe STO dynamics.

\section{Comparison to Nonlinear Auto-Oscillator Theory}

\begin{figure}[ht]
\includegraphics[width=\linewidth]{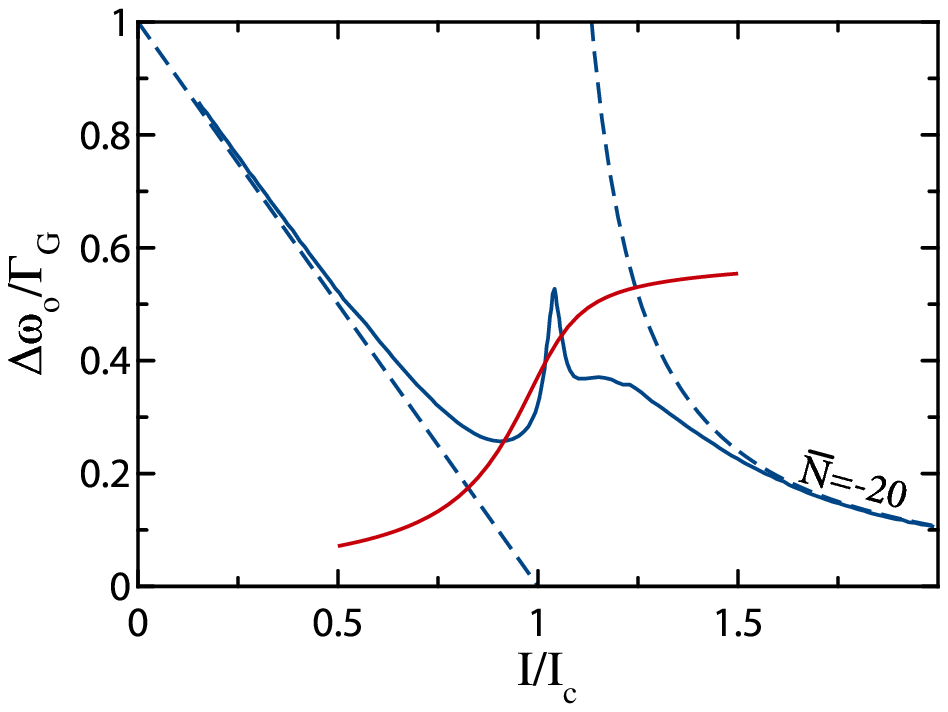}
\caption{\label{fig:KTS-comparison}(Color Online) Comparison of the linewidths predicted by the full KTS auto-oscillator formalism (solid blue lines) along with their asymptotes (dotted blue lines), and the results from applying Eq. (\ref{eqn:spectra2}) (solid red lines). $\bar{N}$ is a measure of the system's nonlinearity, negative values of which indicate a redshift with increasing current $I$. The abscissa is scaled by the critical current $I_c$, while the vertical axis is given in terms of the half-width at half-maximum $\Delta \omega$ and the linear relaxation rate $\Gamma_G$. Data from the full KTS formalism was taken from Fig. 3 of Ref. \citealp{Kim2008a}.}
\end{figure}

The nonlinear auto-oscillator theory developed by Kim, Tiberkevich, and Slavin (KTS) has been quite successful at predicting spectral properties of STOs \cite{Georges2008, Boone2009, Georges2009,Urazhdin2010,Quinsat2012}. The caveat of the KTS approach is that while it predicts simple asymptotic forms for the sub- and super-threshold linewidths, results near threshold can only be calculated by the involved process of finding the eigenfunctions and eigenvalues of the Fokker-Planck equation for the oscillator's power. Our approach, meanwhile, performs best precisely in the realm where the KTS theory becomes computationally burdensome. 

We plot in Fig. \ref{fig:KTS-comparison} the KTS results for the generation linewidth of the magnetic system described in Ref. \citealp{Kim2008a}. We then use the stationary probability distributions (as per Eq. (83b) of Ref. \citealp{Slavin2009a}), dimensionless powers, and frequencies calculated within the KTS framework (ignoring the explicit character of the underlying conservative orbits) as inputs to Eq. (\ref{eqn:spectra2}) in order to judge the agreement with our ensemble-averaged method. There is a slight difference in the plotted quantities: while the KTS linewidths are taken from a single Lorentzian fit to an asymmetric power spectral density, our own results are simply the half-width at half-maximum taken from the distributions calculated using Eq. (\ref{eqn:spectra2}). Our method favorably reproduces the spectral line widths near threshold, confirming its generality.

Combining our expression with the asymptotes of the KTS formalism, one can now easily calculate the expected linewidths of a STO across the entire domain of currents from sub- to super-threshold.

%\bibliography{cleaned-abbrev}